\documentclass[review]{elsarticle}

\usepackage{hyperref,epstopdf}

\journal{ABC}

\usepackage{geometry}
\usepackage{geometry}
\usepackage{bm}   
\usepackage{amsthm}
\usepackage{algorithm}%
\usepackage{algorithmicx}%
\usepackage{amssymb, amsmath, booktabs, color, bbm}
\newtheorem{thm}{Theorem}
\newtheorem{lem}{Lemma}
\newtheorem{prop}{Proposition}
\newtheorem{Example}{Example}
\newtheorem{remark}{Remark}
\newtheorem{assumption}{Assumption}

\newcommand{\E}{\mathbb{E}}

\usepackage{csvsimple}
\usepackage{array}
\usepackage{caption}
\usepackage{etoolbox}
\usepackage{multirow}
\usepackage{float}      

\begin{document}

\begin{frontmatter}

\title{Goodness-of-fit Test for Generalized Functional Linear Models via Projection Averaging}

\author[1]{Feifei Chen}
\author[1]{Kaiming Zhang}
\author[1]{Yanni Zhang}
\author[2]{Hua Liang\footnote{Corresponding author: Hua Liang, hliang@gwu.edu.}}
\address[1]{Faculty of Arts and Sciences, Beijing Normal University, Zhuhai, 519087, China}
\address[2]{Department of Statistics, George Washington University, Washington, D.C. 20052, USA.}

\begin{abstract}
Assessing model adequacy is a crucial step in regression analysis, ensuring the validity of statistical inferences. For Generalized Functional Linear Models (GFLMs), which are widely used for modeling relationships between scalar responses and functional predictors, there is a recognized need for formal goodness-of-fit testing procedures. Current literature on this specific topic remains limited.
This paper introduces a novel goodness-of-fit test for GFLMs. The test statistic is formulated as a $U$-statistic derived from a Cramér-von-Mises metric integrated over all one-dimensional projections of the functional predictor. This projection averaging strategy is designed to effectively mitigate the curse of dimensionality. We establish the asymptotic normality of the test statistic under the null hypothesis and prove the consistency under the alternatives. As the asymptotic variance of the limiting null distribution can be complex for practical use, we also propose practical bootstrap resampling methods for both continuous and discrete responses to compute $p$-values. Simulation studies confirm that the proposed test demonstrates good power performance across various settings, showing advantages over existing methods.
\end{abstract}

\begin{keyword}
Generalized functional linear models\sep Model checking\sep $U$statistic\sep Projection Averaging

\end{keyword}

\end{frontmatter}


\section{Introduction}\label{sec:intr}

\par Recently, works on functional models have been intensively studied due to the popularity of functional data analysis. Among these works, in many cases, we want to model the relationship between a real-valued response and a functional predictor. An early and influential work by \cite{james2002gflms} extended generalized linear models to this context, proposing the Generalized Functional Linear Model (GFLM). A canonical example of this model, detailed in their paper, is found in medical research: predicting a binary outcome, such as five-year survival, based on the trajectory of a patient's bilirubin levels over time.

Consider the following GFLM,
\begin{eqnarray}\label{model}
  \E\left\{ Y|X \right\} = F \left(\alpha_0 + \int_{0}^1 X(t) \beta_0(t)dt \right),
\end{eqnarray}
where $Y$ is a univariate response variable taking values in $\mathcal{Y}$, a subset of real numbers, and its distribution is assumed to be a member of the exponential family. $X(\cdot)$ is a random predictor process that take values in $L^2(\mathbb{I})$, the space of square integrable functions defined on $\mathbb{I}= [0,1]$.
$F$ is a known link function, $\alpha_0$ is an unknown scalar and $\beta_0(\cdot)$ is an unknown infinite dimensional
slope function. The conditional mean with respect to $X=X(\cdot)$ can be understood as a function of a collection of random variables $\{X(t): 0\leq t \leq 1\}$ throughout this paper. 

\par Early estimation methods for model \eqref{model} include the Functional Principal Component Analysis (FPCA) based "reduce-then-regress" strategy by \cite{muller&stadtmuller2005model} and the direct penalized spline approach by \cite{CARDOT200524}. The theoretical properties of the popular FPCA method were then extensively studied. \cite{10.1214/009053606000000957} established its optimal rate in the linear case but also identified its instability with closely spaced eigenvalues, while \cite{dou2012} confirmed the optimal minimax rates for the FPCA-based estimator in the GFLM setting. To address the instability of the FPCA method, \cite{yuan&cai2010flm} introduced a smoothness regularization framework based on Reproducing Kernel Hilbert Spaces (RKHS). This RKHS framework was later generalized to the GFLM setting by \cite{shang&cheng2015inference}. They established a Bahadur representation for functional data, which provides a unified treatment for various inference problems. Based on this Bahadur representation, they constructed asymptotically valid confidence intervals and prediction intervals, and proposed a penalized likelihood ratio test for testing the slope function. We therefore adopt this approach to estimate model \eqref{model} before implementing the testing procedure, as it provides both a robust estimation framework and the necessary theoretical inferential tools for our analysis.


\par When building a regression model, the interpretations could be misleading when the model is not adequate. It is therefore necessary to check the adequacy of a model, which is often performed through a goodness-of-fit testing procedure. It tests the validity of the considered model against general nonparametric alternatives in an omnibus way and can tell whether a more complex model is necessary or not. The primary objective of this paper is to develop a goodness-of-fit test for model~\eqref{model}, which encompasses functional linear regression models, functional logistic regression models, and functional Poisson regression models as special cases. Specifically, we test the following hypothesis in an omnibus way:
\begin{eqnarray}\label{H0}
  H_0: P \left( \E\{Y | X\}= F \left(\alpha_0 + \int_{0}^1 X(t) \beta_0(t)dt\right)\right) = 1, \quad \mbox{for some } (\alpha_0, \beta_0) \in \mathcal{H},
\end{eqnarray}
against
\begin{eqnarray}\label{H1}
  H_1: P \left( \E\{Y | X\}\neq F \left(\alpha_0 + \int_{0}^1 X(t) \beta(t)dt\right) \right) > 0, \quad \mbox{for any } (\alpha, \beta) \in \mathcal{H}.
\end{eqnarray}
Here $\mathcal{H}$ denotes the parameter space of $(\alpha, \beta)$, which will be specified in Section 2.2.

\par There are many well-developed techniques and methods for goodness-of-fit tests for finite-dimensional models.
For instance, one prominent method is the use of residual-marked empirical processes (e.g., \citealp{stute1997nonparametric} and \citealp{stute&zhu1998}), and another is the application of nonparametric estimations (e.g., \citealp{hardle1993comparing} and \citealp{zheng1996consistent}).
As the dimension of the predictors increases, the power performance of most tests greatly deteriorates due to the curse of dimensionality. To alleviate this problem, \cite{Zhu&An1992}, \cite{Zhu&Li1998}, \cite{Escanciano2006} and \cite{Xia2009} used projection-based methods. However, all these techniques are fundamentally designed for finite-dimensional predictors. Their direct application to functional models is therefore infeasible due to the infinite-dimensional nature of the data, which introduces unique complexities to the testing procedure and its theoretical properties.

Most of the existing research on goodness-of-fit tests for functional regression models focuses on linear models.
For the scalar-on-function linear models, where the response is scalar and the predictor is functional, \cite{GarciaPortugues2014} and \cite{CuestaAlbertos2019} constructed goodness-of-fit tests using projected empirical processes. The former integrated all projection directions, but no results on the weak convergence of the test statistic were provided. In contrast, the latter adopted several (not all) randomly chosen projection directions and then aggregated the resulting $p$-values.
After projecting the functional predictor to the least favorable direction for the null hypothesis, \cite{Patilea2020} proposed a test statistic based on nonparametric kernel smoothing methods.
Recently, \cite{Shi2025} developed an adaptive test that combines conventional global smoothing techniques with kernel-based methods. 
When the response is missing at random, \cite{Febrero-Bande2024} proposed a test statistic based on a marked empirical process. 
\cite{Zhang2023} considered a more complex scenario where the model also incorporates scalar predictors measured with additive errors. For this setting, they proposed a kernel-based test statistic constructed from residuals of the null model, which were obtained using the corrected profile least squares estimation.
\cite{Shi2021} and \cite{Xia2024JSSC} studied the adequacy test of scalar-on-function linear quantile regression models. 
When the response is functional, \cite{chen2020model} and \cite{Garcia-Portugues2021} proposed goodness-of-fit tests based on residual marked empirical processes for function-on-scalar and function-on-function linear models, respectively. 
For comprehensive surveys on the topic of goodness-of-fit tests for models involving functional data, one can refer to the reviews by \cite{GonzlezManteiga2022ARO} and \cite{Gonzalez-Manteiga2023review}.

Goodness-of-fit tests for functional generalized linear or nonlinear regression models have received limited scholarly attention. To our knowledge, in the domain of functional single-index models that connect a functional predictor to a scalar response via an unknown link function, \cite{Xia2023} introduced an adaptive-to-model test for a given link function using functional sliced inverse regression. Meanwhile, \cite{Chan2023} developed a specification test with a $U$-statistic to verify the form of the link function. Besides, \cite{Li2024} proposed a generalized functional regression test for the hypothesis $\Pr \big( \E\{ \mathcal{L}(\varepsilon) | X\} =0 \big) = 1$, where $\mathcal{L}(\cdot)$ is a prespecified function of the model error $\varepsilon$. Therefore, a standard model checking test can be viewed as a special case of their framework, occurring when $\mathcal{L}(\varepsilon)$ is simply equal to $\varepsilon$.

In this paper, we propose a novel global test statistic in the form of a $U$-statistic derived from a Cramér-von-Mises metric to address the goodness-of-fit testing problem for GFLMs. Our approach is founded on a projection averaging strategy that systematically captures any model misspecification by projecting the functional predictors onto all possible one-dimensional directions, thereby overcoming the curse of dimensionality inherent in functional data. A key methodological contribution of this work is the transformation of a Cramér-von-Mises metric, which contains a complex integral over infinite-dimensional projection directions, into a computationally feasible $U$-statistic. This formulation not only greatly simplifies the calculation but also provides a solid foundation for the subsequent asymptotic analysis. Theoretically, our main contribution is the rigorous derivation of the asymptotic normality of the $U$-statistic under the null hypothesis.  Furthermore, we establish the consistency of our test under global alternatives. To facilitate its practical application, we also develop bootstrap procedures to compute the $p$-values for both continuous and discrete responses. Extensive simulation studies validate our theoretical findings and demonstrate the preponderance of the proposed method in terms of statistical power.

\par The rest of this paper is organized as follows. In Section~\ref{procedure}, we construct the test procedure and provide a brief outline of the estimation method to make the paper self-contained. Section~\ref{Asymptotics} presents the asymptotic properties of the test statistic under the null and alternative hypotheses. In Section~\ref{NumStu}, bootstrap resampling procedures for both continuous and discrete responses are introduced to determine critical values. Simulation studies are then conducted to examine the performance of the proposed test. 
Section \ref{conclusion} concludes with some discussions. Proofs of the theoretical results are given in the Appendix.

\section{Testing the generalized functional linear model}\label{procedure}
\subsection{The basic idea and test statistic}\label{sec2.1}
\par 
To address the difficulty caused by the infinite dimensionality of the functional predictors, we approximate $X$ with a truncated basis expansion, where the truncated dimension increases asymptotically as the sample size increases. 
Specifically, let $\{\psi_\nu\}_{\nu\ge 1} \subset L^2(\mathbb{I})$ be an arbitrarily fixed orthonormal basis of the function space $L^2(\mathbb{I})$. 
Then $X$ can be expanded as
\begin{eqnarray}\nonumber
  X = \sum_{\nu=1}^\infty \langle X, \psi_\nu \rangle_{L^2} \psi_\nu,
\end{eqnarray}
where $\langle \cdot,\cdot \rangle_{L^2}$ denote the inner product in $L^2(\mathbb{I})$.
That is, $\langle X, \psi_\nu \rangle_{L^2} = \int_{0}^1 X(t) \psi_\nu(t)dt$.
Define
$ X^p = \left(\langle X, \psi_1 \rangle_{L^2}, \ldots, \langle X, \psi_p \rangle_{L^2}\right)^{\top},$
where the superscript $\top$ denotes the transpose. The following proposition enables us to transfer functional predictors to their projections. It extends the results in \cite{Zhu&An1992}, \cite{Zhu&Li1998}, and \cite{Escanciano2006} to the functional predictors case and is similar to Lemma 1 in \cite{patilea2016testing}.

\begin{prop}\label{prop1}
Assume that $e \in\mathbb{R}$ is a random variable with $E[e]=0$. Then, for any stochastic process $X\in L^2(\mathbb{I})$, we have the following equivalence: 
\begin{eqnarray}
  \E\{e | X\} = 0 ~a.s. &\Longleftrightarrow & \E\{e | \gamma^{\top} X^p\} = 0 ~a.s.~ \forall \gamma \in \mathbb{S}^p, ~\forall p\geq 1, \nonumber \\
   &\Longleftrightarrow&  \E\left\{e I\left(\gamma^{\top} X^p \leq x \right) \right\} = 0 ~a.s.~ \forall x\in \mathbb{R}, ~\forall \gamma \in \mathbb{S}^p, ~\forall p\geq 1,  \label{projemp}
\end{eqnarray}
where $I(\cdot)$ denotes the indicator function, $\mathbb{S}^p=\{\gamma \in \mathbb{R}^p: \|\gamma\|=1\}$ is the unit sphere in $\mathbb{R}^p$, and $\|\cdot\|$ denotes the Euclidean norm.
\end{prop}

Let $e = Y - F \big(\alpha_0 + \int_{0}^1 X(t) \beta_0(t)dt \big)$ denote the error term in model \eqref{model}, then the null hypothesis in \eqref{H0} is equivalent to $\E\{e|X\}=0, a.s.$. 
Proposition~\ref{prop1} implies that a test for $H_0$ can be constructed based on one-dimensional projections. This motivates us to consider test statistics based on the deviation between $E\{e I\left(\gamma^{\top} X^p \leq x \right)\}$ and zero.
Consider $\gamma$ to be a random vector and $\mu(\cdot)$ is the distribution function of $\gamma$. Let $F_{\gamma}(\cdot)$ be the distribution function of $\gamma^{\top} X^p$ for a given $\gamma$ and $p$. Then a Cramér-von-Mises type norm of $E\{e I\left(\gamma^{\top} X^p \leq x \right)\}$ can be given by
\begin{eqnarray*}
  && \int_{\mathbb{S}^p} \int_\mathbb{R} \Big\{ \E\left\{e I\left(\gamma^{\top} X^p \leq x \right) \right\} \Big\}^2 dF_{\gamma}(x)d\mu(\gamma) \\
  &=& \int_{\mathbb{S}^p} \int_\mathbb{R} \E\left\{e_1 I\left(\gamma^{\top} X_1^p \leq x \right) e_2 I\left(\gamma^{\top} X_2^p \leq x \right) \right\} dF_{\gamma}(x)d\mu(\gamma)   \\
  &=& \E\left\{ e_1e_2 \int_{\mathbb{S}^p} I\left(\gamma^{\top} X_1^p \leq \gamma^{\top} X_3^p \right) I\left(\gamma^{\top} X_2^p \leq \gamma^{\top} X_3^p \right) d\mu(\gamma) \right\},
\end{eqnarray*}
where $e_1, e_2$ are i.i.d. copies of $e$, and $X_1, X_2, X_3$ are i.i.d. copies of $X$, respectively. 
Further assume that $\gamma$ is uniformly distributed on $\mathbb{S}^p$. Then, the following lemma, which was first proved by \cite{Escanciano2006} and recently generalized by \cite{Kim2020}, provides an explicit form for the integration.
\begin{lem}\label{lem:angle_integral}
For any nonzero vectors $U_1, U_2 \in \mathbb{R}^p$, we have
$$
\int_{\mathbb{S}^p} I(\gamma^{\top}U_1 \leq 0) I(\gamma^{\top}U_2 \leq 0) d\mu(\gamma) = \frac{1}{2} - \frac{1}{2\pi}\mathrm{Ang}(U_1, U_2),
$$
where $\mathrm{Ang}(U_1, U_2) = \arccos\left(U_1^{\top}U_2/\{\|U_1\|\|U_2\|\}\right)$ is the angle between the vectors $U_1$ and $U_2$.
\end{lem}

According to Lemma~\ref{lem:angle_integral}, we have
\begin{eqnarray*}
\int_{\mathbb{S}^p} I\left(\gamma^{\top} X_1^p \leq \gamma^{\top} X_3^p \right) I\left(\gamma^{\top} X_2^p \leq \gamma^{\top} X_3^p \right) d\mu(\gamma)
&=& \frac{1}{2} - \frac{1}{2\pi}\mathrm{Ang}\left(X_1^p - X_3^p, X_2^p - X_3^p\right) \\
&=& \frac{1}{2\pi} \mathrm{Ang}\left(X_1^p-X_3^p, X_3^p-X_2^p\right).
\end{eqnarray*}
The final step holds due to the geometric property that for any two vectors $U_1$ and $U_2$, the angle $\mathrm{Ang}(U_1, -U_2) = \pi - \mathrm{Ang}(U_1, U_2)$.
For simplicity, we denote $\mathrm{Ang}\left(X_1^p-X_3^p, X_3^p-X_2^p\right)$ as $\mathrm{Ang}\left(X_{13}^p, X_{32}^p\right)$ in the following. Then the Cramér-von-Mises type norm of $\E\{e I\left(\gamma^{\top} X^p \leq x \right)\}$ can be written as
\begin{eqnarray*}
  \int_{\mathbb{S}^p} \int_\mathbb{R} \Big\{ \E\left\{e I\left(\gamma^{\top} X^p \leq x \right) \right\} \Big\}^2 dF_{\gamma}(x)d\mu(\gamma)
      = \frac{1}{2\pi}\E\big\{ e_1e_2 \mathrm{Ang}\left(X_{13}^p, X_{32}^p\right) \big\}.
\end{eqnarray*}

According to the equivalence presented in Proposition \ref{prop1}, we can construct the test statistic based on an empirical version of $\E\left\{ e_1e_2 \mathrm{Ang}(X_{13}^p, X_{32}^p) \right\}$. 
Recall that $ X^p = \left(\langle X, \psi_1 \rangle_{L^2}, \ldots, \langle X, \psi_p \rangle_{L^2}\right)^{\top}$, which depends on the orthonormal basis $\{\psi_\nu\}_{\nu\ge 1} \subset L^2(\mathbb{I})$.
Clearly, the practitioner would prefer a basis that allows
for an accurate low-dimensional approximation of $X$ and hence allows for a low $p$ in the test procedure. 
To this end, we employ functional principal components, which provide optimal low-dimensional representations of $X$ with respect to the mean squared error. Specifically, we first estimate the covariance operator of $X$ from the sample curves $X_i$, $i=1,\ldots,n$, and then perform an eigen-decomposition to obtain the estimated eigenfunctions $\hat\psi_\nu$, $\nu=1,\ldots,p$. Consequently, the functional data $X_i$ are approximated by $ \widehat X_i^p = \big(\langle X_i, \hat\psi_1 \rangle_{L^2}, \ldots, \langle X_i, \hat\psi_p \rangle_{L^2}\big)^{\top}$. 
Furthermore, a $U$-statistic type test statistic is defined as
\begin{eqnarray}\label{test_Tn} 
  T_n
  = \frac{1}{n(n-1)(n-2)} \sum_{i\neq j\neq k}^n \hat{e}_i\hat{e}_j \mathrm{Ang}\left( \widehat X_{ik}^p, \widehat X_{kj}^p\right),
\end{eqnarray}
where $(X_i,Y_i), i=1,\ldots,n,$ are i.i.d. copies of $(X,Y)$ and $\hat{e}_i=Y_i-F(\hat{\alpha}_{n} + \int_0^1 X_i(t)\hat{\beta}_{n}(t) dt), i=1,\ldots,n$, are residuals.
Here $\hat{\alpha}_{n}$ and $\hat{\beta}_{n}$ denote the estimators of parameters $\alpha$ and $\beta$ under $H_0$, respectively.
The estimation procedures of the unknown parameters will be discussed in the following subsection.

\subsection{Estimation of parameters} \label{EstProc}

The estimators of the intercept $\alpha_0$ and the slope function $\beta_0$ play an important role in deriving the asymptotic properties of the test statistic $T_n$ in \eqref{test_Tn}.
We adopt a roughness regularization approach proposed by \cite{shang&cheng2015inference} to estimate $\alpha_0$ and $\beta_0$ in an RKHS framework. They presented the convergence rate of estimators and established a Bahadur representation for functional data, which are essential to our theoretical analyses. To make the paper self-contained, we provide a brief introduction to the estimation procedures and notations below.

Let $\beta\in H^m(\mathbb{I})$, the $m$-order Sobolev space defined by
\begin{align*}
H^m(\mathbb{I}) = \{&\beta: \mathbb{I} \mapsto \mathbb{R} \mid \beta^{(j)}, j = 0, \ldots, m-1, \text{are absolutely continuous, and } \beta^{(m)} \in L^2(\mathbb{I})\}.
\end{align*}
Therefore, the unknown parameter $\theta=(\alpha, \beta)$ belongs to $\mathcal{H}= \mathbb{R}\times H^m(\mathbb{I}).$ 
Based on the i.i.d. samples $(X_i, Y_i), i=1,\ldots,n,$, the regularized estimator $\hat{\theta}_n = (\hat{\alpha}_{n}, \hat{\beta}_{n})$ is given by
\begin{align}\label{eq:estimator_def}
    (\hat{\alpha}_{n}, \hat{\beta}_{n}) 
    &= \sup_{(\alpha,\beta)\in\mathcal{H}} \left\{ \ell_{n,\lambda}(\theta) \right\} \nonumber\\
    &= \sup_{(\alpha,\beta)\in\mathcal{H}} \left\{ \frac{1}{n}\sum_{i=1}^n \ell\left(Y_i; \alpha + \int_0^1 X_i(t)\beta(t)dt\right) - \frac{\lambda}{2} J(\beta, \beta) \right\},
\end{align}
where $\ell(y;a)$ is the log-likelihood 
function defined over $y\in \mathcal{Y}$ and $a\in \mathbb{R}$, $\lambda \ge 0$ is a non-negative tuning parameter, $J(\beta, \tilde{\beta}) = \int_0^1 \beta^{(m)}(t)\tilde{\beta}^{(m)}(t) dt$ is a a roughness penalty.
Note that we suppress the $\lambda$ in $(\hat{\alpha}_{n}, \hat{\beta}_{n})$ for simplicity.
The smoothness and tail conditions on $\ell$ are given in Assumption \ref{assump:model} below.
\begin{assumption} \label{assump:model}
\noindent (a) $\ell(y; a)$ is three times continuously differentiable and strictly concave w.r.t. $a$. There exist positive constants $C_0$ and $C_1$ s.t.,
\begin{equation*} 
\begin{split}
    \E\left\{\exp\left(\sup_{a \in \mathbb{R}}|\ddot{\ell}_a(Y; a)|/C_0\right) ~\Big|~ X\right\} &\le C_1, \\
    \E\left\{\exp\left(\sup_{a \in \mathbb{R}}|\ell_a'''(Y; a) /C_0\right) ~\Big|~ X\right\} &\le C_1, \quad \text{a.s.,}
\end{split}
\end{equation*}
where $\ddot{\ell}_a(y;a)$ and $\ell_a'''(y;a)$ are the second- and third-order derivatives of $\ell(y;a)$ w.r.t. $a$.

\noindent (b) There exists a positive constant $C_2$ s.t.,
\[
C_2^{-1} \le B(X) \equiv -\E\left\{\ddot{\ell}_a\left(Y; \alpha_0 + \int_0^1 X(t)\beta_0(t) dt\right) ~\Big|~ X\right\} \le C_2, \quad \text{a.s.}
\]
In addition, $X$ is weighted-centered in the sense that $E\{B(X)X(t)\} = 0$ for any $t \in \mathbb{I}$.

\noindent (c) $\epsilon \equiv \dot{\ell}_a\left(Y; \alpha_0 + \int_0^1 X(t)\beta_0(t) dt\right)$ satisfies $\E\{\epsilon|X\} = 0$ and $\E\{\epsilon^2|X\} = B(X)$, a.s., where $\dot{\ell}_a(y;a)$ is the first-order derivative of $\ell_(y;a)$ w.r.t. $a$.
\end{assumption}

\begin{remark}\label{remark:error}
Note that $\epsilon = \dot{\ell}_a\left(Y; \alpha_0 + \int_0^1 X(t)\beta_0(t) dt\right)$ in Assumption \ref{assump:model}(c) is typically called \textit{score} function in the generlized linear model framework. It is easy to calculate that
\begin{align*}
   \epsilon = \frac{Y-\E\{Y|X\}}{a(\phi)},
\end{align*}
where $\phi$ is a possible nuisance parameter in the exponential family and $a(\cdot)$ is a known function. $\phi$ is typically assumed to be identical over all observations, such as the variance $\sigma^2$ in linear regression \cite[see][\S 5.2]{Hardle2004}.
Then we can conclude that, under the null hypothesis in \eqref{H0}, the error term 
\begin{align*}
    e_i = Y_i - F \left(\alpha_0 + \int_{0}^1 X_i(t) \beta_0(t)dt \right) = a(\phi)\epsilon_i, \quad i=1\ldots,n, \quad \text{a.s.,} 
\end{align*} 
and 
\begin{align*} 
    \E\{e^2 | X \} = a^2(\phi) \E\{ \epsilon^2 | X \} = a^2(\phi) B(X). 
\end{align*} 
\end{remark}

To analyze the asymptotic properties of the estimator in \eqref{eq:estimator_def} within a RKHS framework, \cite{shang&cheng2015inference} defined the inner product in $H^m(\mathbb{I})$ based on the weighted covariance function $C(s, t) = \E\{B(X)X(t)X(s)\}$, where the weighting term $B(X)$ is related to the link function and is defined in Assumption \ref{assump:model}.
Specifically, for any $\beta, \tilde{\beta} \in H^m(\mathbb{I})$, the inner product in $H^m(\mathbb{I})$ is defined as
\begin{align*}
   \big\langle \beta, \tilde{\beta} \big\rangle_{H^m} 
      = \int_0^1 \int_0^1 C(s, t)\beta(t)\tilde{\beta}(s) ds dt + \lambda J(\beta, \tilde{\beta}) 
      =: V(\beta, \tilde{\beta}) + \lambda J(\beta, \tilde{\beta}).
\end{align*}
\cite{shang&cheng2015inference} showed that $H^m(\mathbb{I})$ is indeed a RKHS under $\langle \cdot, \cdot \rangle_{H^m}$ if the following Assumption \ref{assump:covariance_op} holds.
Furthermore, for any $\theta = (\alpha, \beta)$, $\tilde{\theta} = (\tilde{\alpha}, \tilde{\beta}) \in \mathcal{H}$, define
\begin{align*}
   \big\langle \theta, \tilde{\theta} \big\rangle_{\mathcal{H}} 
   &= \E\left\{B(X)\left(\alpha + \int_0^1 X(t) \beta(t)\,dt\right) \left(\tilde{\alpha} + \int_0^1 X(t)\tilde{\beta}(t)\,dt\right)
   \right\}  + \lambda J(\tilde{\beta}, \beta) \\
   &= \E\{B(X)\} \alpha\tilde{\alpha} + \big\langle \beta, \tilde{\beta} \big\rangle_{H^m}.
\end{align*}
By Proposition 2.1 in \cite{shang&cheng2015inference}, $\mathcal{H}$ is a Hilbert space under $\langle \cdot, \cdot \rangle_{\mathcal{H}}$.
We denote the corresponding norms of $\langle \cdot, \cdot \rangle_{H^m}$ and $\langle \cdot, \cdot \rangle_{\mathcal{H}}$ by $\|\cdot\|_{H^m}$ and $\|\cdot\|_{\mathcal{H}}$, respectively.

\begin{assumption} \label{assump:covariance_op}
$C(s, t)$ is continuous on $\mathbb{I} \times \mathbb{I}$. Furthermore, define a linear bounded operator $\mathcal{C}(\cdot)$ from $L^2(\mathbb{I})$ to $L^2(\mathbb{I}): (\mathcal{C}\beta)(t) = \int_0^1 C(s,t)\beta(s)\,ds$, for any $\beta \in L^2(\mathbb{I})$ satisfying $\mathcal{C}\beta = 0$, we have $\beta = 0$.
\end{assumption}

As pointed out by \cite{shang&cheng2015inference}, a cornerstone of the theoretical analysis is the construction of an eigen-system in $H^m(\mathbb{I})$ that simultaneously diagonalizes both $V(\cdot, \cdot)$ and $J(\cdot, \cdot)$. 
We assume that such an eigen-system exists; see Sections S.2--S.5 in the supplement to \cite{shang&cheng2015inference} for more details about how to construct an eigen-system satisfying the following Assumption \ref{assump:eigensystem}.   

\begin{assumption} \label{assump:eigensystem}
There exists a sequence of functions $\{\varphi_\nu\}_{\nu\ge 1} \subset H^m(\mathbb{I})$ such that $\|\varphi_\nu\|_{L^2} \le C_\varphi \nu^a$ for each $\nu \ge 1$, some constants $a \ge 0, C_\varphi > 0$ and
$$
V(\varphi_\nu, \varphi_\mu) = \delta_{\nu\mu}, \quad J(\varphi_\nu, \varphi_\mu) = \rho_\nu\delta_{\nu\mu} \quad \text{for any } \nu, \mu \ge 1,
$$
where $\delta_{\nu\mu}$ is Kronecker's notation, and $\rho_\nu$ is a nondecreasing nonnegative sequence satisfying $\rho_\nu \asymp \nu^{2k}$ for some constant $k > a + 1/2$. 
Here we denote $a_n \asymp b_n$ if and only if there exist positive constants $c_1$, $c_2$ such that $c_1 \leq a_\nu/b_\nu \leq c_2 $ for all $\nu$.
Furthermore, any $\beta \in H^m(\mathbb{I})$ admits the Fourier expansion $\beta = \sum_{\nu=1}^{\infty} V(\beta, \varphi_\nu)\varphi_\nu$.
\end{assumption}




We present the regularity conditions on the predictor process $X$ and rate conditions on the penalty parameter $\lambda$ to obtain the convergence rate and the Bahadur representation for GFLMs in Assumptions \ref{assump:moment_cond} and \ref{assump:rate}, respectively.

\begin{assumption} \label{assump:moment_cond}
Let $\|X\|_{L^2}^2 = \int_0^1 X^2(t) dt$. There exists a constant $s \in (0, 1)$ such that
\[
\E\big\{\exp(s\|X\|_{L^2})\big\} < \infty.
\]
Moreover, suppose that there exists a constant $M_0 > 0$ such that for any $\beta \in H^m(\mathbb{I})$,
\[
\E\big\{ \langle X, \beta \rangle_{L^2}^4 \big\}  \le M_0 \left[ \E\big\{\langle X, \beta \rangle_{L^2}^2 \big\} \right]^2.
\]
\end{assumption}

\begin{assumption} \label{assump:rate}
Denote $h = \lambda^{1/(2k)}$, where $k$ is specified in Assumption \ref{assump:eigensystem}. As $n \to \infty$, 
$h = o(1)$, $n^{-1/2}h^{-1} = o(1)$, $\log(h^{-1}) = O(\log n)$, and $n^{-1/2}h^{-(a+1)-((2k-2a-1)/(4m))}(\log n)^2(\log\log n)^{1/2} = o(1)$ hold, where $a$ is also specified in Assumption \ref{assump:eigensystem}.
\end{assumption}


The following lemmas, adapted from \cite{shang&cheng2015inference}, present the convergence rate and the Bahadur Representation of $\hat{\theta}_n$, respectively.

\begin{lem}[Convergence rate] \label{lem:rate_satisfied}
Suppose that Assumptions \ref{assump:model}--\ref{assump:rate} hold, then
$$\|\hat{\theta}_{n} - \theta_0\|_{\mathcal{H}} = O_P(r_n),$$
where $r_n = (nh)^{-1/2} + h^k$. 
\end{lem}

\begin{lem}[Bahadur Representation]\label{lem:bahadur}
Suppose that Assumptions \ref{assump:model}--\ref{assump:rate} hold. Then
\[
\|\hat{\theta}_{n} - \theta_0 - S_{n}(\theta_0)\|_{\mathcal{H}} = O_P(a_n),
\]
where $S_n(\cdot)$ denotes the Fr\'{e}chet derivative of the criterion function $\ell_{n,\lambda}(\cdot)$ in \eqref{eq:estimator_def} with respect to $\theta$.
Here we suppress the $\lambda$ in $S_n(\cdot)$ for simplicity.
The remainder rate $a_n$ is given by
\[
a_n = n^{-1/2}h^{-(4ma+6m-1)/(4m)}r_n(\log n)^2(\log\log n)^{1/2} + C_\ell h^{-1/2}r_n^2,
\]
with $C_\ell \equiv \sup_{x \in L^2(\mathbb{I})} \E\left\{\sup_{a \in \mathbb{R}} |\ell_a'''(Y;a)| \mid X=x\right\}$.
\end{lem}

\section{Asymptotic properties}\label{Asymptotics}

\par In this section, we investigate the asymptotic behaviors of $T_n$ under the null hypothesis and the alternatives, respectively. 
The following theorem states the asymptotic normality of the test statistic $T_n$ under the null hypothesis.


\begin{thm}\label{thm:H0}
Suppose that Assumptions \ref{assump:model}--\ref{assump:rate} hold, the data $(X_i,Y_i), i=1,\ldots,n$, are i.i.d. copies of $(X,Y)$. Under $H_0$ in \eqref{H0}, if $p=o(n^{1/4})$, we have
\begin{align*}
   \frac{n {T}_{n}}{10\sqrt{ 2\E \big\{ H_n^2(Z_1,Z_2) \big\}} }
   \stackrel{d}{\longrightarrow} N(0, 1),
\end{align*} 
where $H_n(Z_1,Z_2)$ is defined in equation $(30)$ and $Z_1$, $Z_2$ are i.i.d. copies of $Z=(e, X)$.
\end{thm}

\begin{remark}\label{remark:H0}
Note that $\E \big\{ H_n^2(Z_1,Z_2) \big\} = O(h^{-2})$ by equation $(31)$, Theorem \ref{thm:H0} indicates that the convergence rate of $T_n$ is $O(nh)$, which is slightly slower than the parametric rate $O(n)$.
This is attributed to the slightly slower convergence rate of $\hat{\theta}_n$, which can be seen in Lemma \ref{lem:rate_satisfied}.
As pointed out by \cite{shang&cheng2015inference}, this is a price to pay for obtaining the Bahadur Representation within the GFLM framework, which is essential to establish the asymptotic normality of $T_n$. 
Besides, although the asymptotic normality has been established, its asymptotic variance, as shown in $(31)$, is rather tricky to estimate. 
Therefore, we employ two different resampling procedures, one for continuous responses and the other for discrete responses, to determine the critical values in numerical studies.
\end{remark}


In the following, we investigate the asymptotic behavior of $T_n$ under the alternative $H_1$ in \eqref{H1}. 
Note that model \eqref{model} is misspecified under $H_1$, while the unknown parameter is still obtained by \eqref{eq:estimator_def}. Denote
\begin{align*}
     (\tilde{\alpha}_0, \tilde{\beta}_0) 
    = \sup_{(\alpha,\beta)\in\mathcal{H}} \E\left\{ \ell\left(Y; \alpha + \int_0^1 X(t) \beta(t) dt\right) \right\},
\end{align*}
and 
\begin{align*}
     \tilde{\epsilon} = \dot{\ell}_a\left(Y; \tilde{\alpha}_0 + \int_0^1 X(t)\tilde{\beta}_0(t) dt\right).
\end{align*}
The following theorem presents the asymptotic behavior of the test statistic $T_n$ under $H_1$.

\begin{thm}\label{thm:H1}
Suppose that Assumptions \ref{assump:model}--\ref{assump:rate} hold, the data $(X_i,Y_i), i=1,\ldots,n$, are i.i.d. copies of $(X,Y)$. Under $H_1$ in \eqref{H1}, if $p=o(n^{1/4})$, we have
\begin{align*}
   T_n \stackrel{p}{\longrightarrow} \E\left\{ \tilde{\epsilon}_1\tilde{\epsilon}_2 \mathrm{Ang}\left(X_{13}, X_{32}\right) \right\} > 0,
\end{align*} 
where $\tilde{\epsilon}_1$ and $\tilde{\epsilon}_2$ are i.i.d. copies of $\tilde{\epsilon}$, and
$$
\mathrm{Ang}\left(X_{13}, X_{32}\right) = \arccos \left( \frac{\left\langle X_1-X_3, X_3-X_2\right\rangle_{L^2}}{\left\|X_1-X_3\right\|_{L^2} \left\|X_3-X_2\right\|_{L^2}} \right).
$$
\end{thm}

Theorem \ref{thm:H1} shows that our proposed test is consistent since $nh T_n \stackrel{p}{\longrightarrow} \infty$ under $H_1$.

\section{Numerical studies}\label{NumStu}

\par We present the numerical performance of the proposed test in this section. The corresponding R code is available.

\subsection{Resampling procedures} \label{Resam}
As the asymptotic variance of the test statistic $T_n$ under the null hypothesis is rather difficult to estimate, we employ different bootstrap resampling procedures to determine the critical values for both continuous and discrete responses.

For a continuous response $Y$, we adopt the wild bootstrap strategy, which is widely used in the framework of model checking. For example, \cite{GarciaPortugues2014} and \cite{CuestaAlbertos2019} used this strategy to calibrate residuals of functional linear models. The following Algorithm \ref{alg:resampling} states the detailed resampling steps.
When the outcome is discrete, the residuals are also discrete, so the wild bootstrap resampling is not applicable.
Motivated by \cite{Li2022check}, we propose a model-based bootstrap strategy to deal with discrete responses. Take the binary response as an example; the following Algorithm \ref{alg:resampling_binary} outlines the detailed steps for functional logistic regression models.

\begin{algorithm}[H]
\caption{Bootstrap resampling procedure for continuous response.}
\label{alg:resampling}
\vspace{0.3em}
\begin{algorithmic}
\State{\bf Step 1.} Compute the test statistic $T_n$ as in \eqref{test_Tn}.

\State{\bf Step 2.} Generate random variables $v_1,\ldots,v_n$ independently from a two-point distribution which takes values $(1\mp \sqrt{5})/2$ with the probabilities $(5\pm \sqrt{5})/10$. Let
  \begin{eqnarray*}
    Y_i^* = F \left(\hat\alpha_n + \int_{0}^1 X_i(t) \hat\beta_n(t)dt \right) + \hat{e}_i v_i, \quad i = 1,\ldots,n.
  \end{eqnarray*}

\State{\bf Step 3.} Based on the bootstrap samples $(X_i, Y_i^*)$, $i=1,\ldots,n$, compute the bootstrap test statistic $T_n^*$ as in \eqref{test_Tn}.

\State{\bf Step 4.} Repeat Steps 2--3 $B$ times to obtain $T_n^{*1}, T_n^{*2},\ldots, T_n^{*B}$ for some large integer $B$.

\State{\bf Step 5.} For a given significance level $\alpha\in(0,1)$, take the $(1-\alpha)$ quantile of $T_n^{*j}$, $j=1,\ldots,B$, as the critical value $c_{n,\alpha}$, and $B^{-1}\sum_{j=1}^B \mathbb{I} \{ T_n^{*j} \geq T_n \}$ as the estimated $p$-value $\hat{p}$.

\State{\bf Step 6.} Reject $H_0$ if $T_n > c_{n,\alpha}$ or $\hat{p} < \alpha$.
\end{algorithmic}
\end{algorithm}

\begin{algorithm}[H]
\caption{Bootstrap resampling procedure for binary response.}
\label{alg:resampling_binary}
\vspace{0.3em}
\begin{algorithmic}
\State{\bf Step 1.} Compute the test statistic $T_n$ as in (\ref{test_Tn}).

\State{\bf Step 2.} Denote $\hat{Y}_i = F (\hat\alpha_n + \int_{0}^1 X_i(t) \hat\beta_n(t)dt)$, $i=1,\ldots,n$.
Let $p_1^*,\ldots,p_n^*$ be $n$ observations sampled from $\hat{Y}_1,\ldots,\hat{Y}_n$ without replacement. Generate $Y_i^*$,$i=1,\ldots,n$, from independent Bernoulli distributions, where $Y_i^*$ has the probability of success $p_i^*$.

\State{\bf Step 3.} Based on the bootstrap samples $(X_i, Y_i^*)$, $i=1,\ldots,n$, compute the bootstrap test statistic $T_n^*$ as in \eqref{test_Tn}.

\State{\bf Step 4.} Repeat Steps 2--3 $B$ times to obtain $T_n^{*1}, T_n^{*2},\ldots, T_n^{*B}$ for some large integer $B$.

\State{\bf Step 5.} For a given significance level $\alpha\in(0,1)$, take the $(1-\alpha)$ quantile of $T_n^{*j}$, $j=1,\ldots,B$, as the critical value $c_{n,\alpha}$, and $B^{-1}\sum_{j=1}^B \mathbb{I} \{ T_n^{*j} \geq T_n \}$ as the estimated $p$-value $\hat{p}$.

\State{\bf Step 6.} Reject $H_0$ if $T_n > c_{n,\alpha}$ or $\hat{p} < \alpha$.
\end{algorithmic}
\end{algorithm}

\subsection{Simulations}\label{Simu} 
\par Our simulation study considers two settings. Example \ref{ex1} presents a functional linear model, while Example \ref{ex2} focuses on a generalized functional linear model with the logit link function. For each setting, all experiments were repeated $500$ times to assess the Monte Carlo performance of the tests. The number of bootstrap replications used to determine the critical value was set to $B = 1000$. The tests are conducted at sample size $n \in \{50, ~100\}$ and significance level $\alpha \in \{1\%, ~5\%, ~10\%\}$. For the number of orthonormal bases $p$, we adopted two distinct strategies: a data-driven approach and a fixed-value approach with $p \in \{5, ~10 \}$. The data-driven $p$ was selected as the minimum number of components required to capture no less than $95\%$ of the total explained variance, whereas the fixed values were employed for comparative purposes. 
All the parameters that need to be set during the estimation process are in accordance with settings of \cite{shang&cheng2015inference}.

\begin{Example}\label{ex1}
In this example, we consider the goodness-of-fit test for the functional linear regression model:
$$
Y_i = \int_{0}^{1} X_i (t) \beta (t) \, dt ~+~ a \int_{0}^{1} X_i (t) X_i (t) \, dt ~+~ \epsilon_i, ~~~~~i=1, \cdots, n
$$
where
\begin{itemize}
    \item  \( X_i (t) =  \sum_{j=1}^{100} \sqrt{\kappa_j} \eta_j \phi_j (t) \). Here the covariance operator of covariate processes has eigenvalues $\kappa_j = j^{-1.7}$ and eigenfunctions $\phi_1 (t) = 1$, $\phi_j (t) = \sqrt{2} \cos((j-1) \pi t)$, $j \geq 2$. The $\eta_j$'s are independent standard normal while $t$ are $1000$ points evenly spaced over $[0,~1]$;
    \item  \( \beta (t) = r \sum_{j=1}^{100} \theta_j \phi_j (t) \). Let $r^2 = 1.5$ and $\theta_j = \bar{\theta}_j / || \bar{\theta} ||_2 $, where $\bar{\theta}_j = b_j \cdot I_j$ for $j = 1, 2$ and $\bar{\theta}_j = 0$ for $j > 2$. Specifically, $b_1, b_2 \stackrel{\text{i.i.d.}}{\sim} Unif(0,1)$, and  $(I_1,I_2)$ follows a multinomial distribution $Mult(1;0.5,0.5)$;
    \item  \( \epsilon_i \) are i.i.d. standard normal error term independent of \( X \);
    \item Take $a \in \{ 0, 0.05,0.10,0.15,0.20\}$ as the distance away from the null. 
\end{itemize}
\end{Example}

\begin{Example}\label{ex2}
In this example, we consider the goodness-of-fit test for the functional logistic regression model:
$$
P(Y=1|X) = \frac{\exp\left(\int_{0}^{1} X (t) \beta (t) \, dt + a \exp \left(\int_{0}^{1} X (t) \beta (t) \, dt\right)\right)} {1 + \exp\left(\int_{0}^{1} X (t) \beta (t) \, dt + a \exp\left(\int_{0}^{1} X (t) \beta (t) \, dt\right)\right)}
$$
where
\begin{itemize}
    \item  \( X_i (t) =  \sum_{j=1}^{100} \sqrt{\lambda_j} \eta_{ij} V_j (t) \). Let $\lambda_j = (j-0.5)^{-2} \pi^{-2}$, $V_j (t) = \sqrt{2} \sin((j-0.5) \pi t)$, for $j =1, 2, \cdots, 100$. The $\eta_{ij}$'s are independent truncated normals, i.e. $\eta_{ij} = \xi_{ij} I_{\{ |\xi_{ij}| \leq 0.5 \}} + 0.5I_{\{ \xi_{ij} > 0.5 \}} - 0.5I_{\{ \xi_{ij} < -0.5 \}}$, and $\xi_{ij}$'s are standard normal random variables. Similarly, $t$ are $1000$ points evenly spaced over $[0,~1]$;
    \item  \( \beta (t) = 3 \times 10^5 ~(t^{11} (1-t)^6) \);
    \item Take $a \in \{ 0, 0.25, 0.50, 0.75, 1\}$ as the distance away from the null. 
\end{itemize}
\end{Example}

\par
\begin{figure}[H]
\centering
\includegraphics[angle=-90, width=1.0\textwidth]{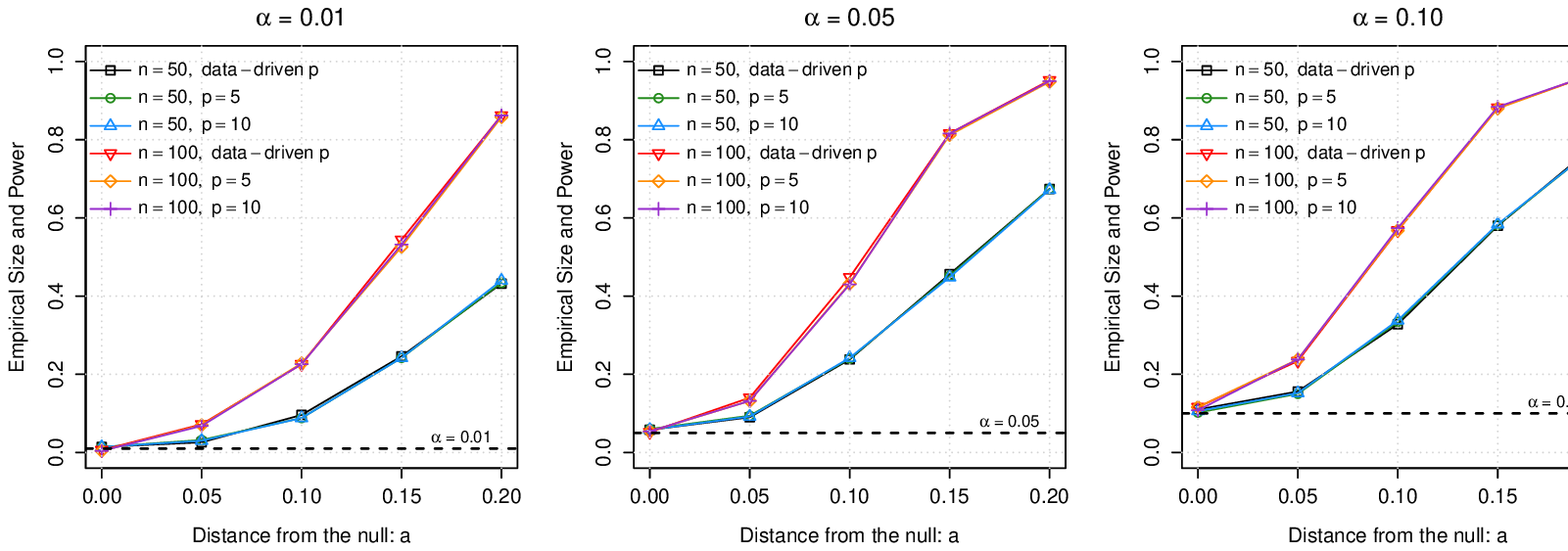}
\caption{Simulation Results for Example \ref{ex1}. The test performance of the proposed test $T_n$, comparing data-driven $p$ and fixed $p = 5, 10$ selection of the orthonormal bases for sample sizes $n = 50, 100$ at 1\%, 5\% and 10\% significance level.}
\label{fig:ex1-1}
\end{figure}

 We first investigate the influence of the number of orthonormal bases $p$ on the test performance. Table \ref{tab:1} reports the results of our proposed test for Example \ref{ex1}. For clearer interpretation, Figure \ref{fig:ex1-1} displays the results at the 5\% significance level from Table \ref{tab:1}. As shown in these results, our test maintains good control of the empirical size and achieves comparable performance across different $p$. 
We adopt the data-driven $p$ for all subsequent simulations.

We next evaluate the finite-sample performance of the proposed test through a comparative simulation study with two methods: the tests presented in \cite{GarciaPortugues2014} and \cite{CuestaAlbertos2019}, which can be implemented via the R functions \texttt{rp.flm.statistic()} and \texttt{PCvM.statistic()}, available in the \texttt{fda.usc} package, respectively. 
Both tests considered the goodness-of-fit test for the functional linear model based on projected empirical processes. The former integrated all projection directions and then defined a statistic that is essentially the average of projected Cra\'{m}er-von Mises statistics (denoted as $PCvM$). While the latter adopted several (not all) randomly chosen projection directions and then constructed both Cra\'{m}er-von Mises type (denoted as $CA_1$) and Kolmogorov-Smirnov type (denoted as $CA_2$) test statistics.
However, once we obtain the model residuals by \cite{shang&cheng2015inference}'s estimation method, we can directly apply the residuals to tests $PCvM$, $CA_1$, and $CA_2$, and obtain the $p$-values by Algorithms \ref{alg:resampling} and \ref{alg:resampling_binary}. For a more comprehensive comparison, we also make comparisons with tests $PCvM$, $CA_1$, and $CA_2$ in Example \ref{ex2}, which involves the functional logistic regression model.
It is also noteworthy that both \cite{GarciaPortugues2014} and \cite{CuestaAlbertos2019} employed the estimation method proposed by \cite{Cardot2007CLT}. To ensure a fair comparison, we report results using both the estimators of \cite{shang&cheng2015inference} and \cite{Cardot2007CLT} in Example \ref{ex1}.

Table \ref{tab:2} reports the empirical size and power of the tests $T_n$, $PCvM$, $CA_1$, and $CA_2$ at various significance levels for Example \ref{ex1}, using the estimators from both \cite{shang&cheng2015inference} (denoted by SC) and \cite{Cardot2007CLT} (denoted by CAP). It can be observed that the influence of the two estimation methods on the test results under the linear regression model is small. 
Note that \cite{Cardot2007CLT} derived a central limit theorem in the functional linear regression model. We conjecture that our proposed test can also be constructed based on  \cite{Cardot2007CLT}'s estimators within a linear framework. 
Furthermore, it can be extended to more complex models as long as a Bahadur representation is obtained.
Irrespective of the estimation method, our test attains higher power than the competing tests. This stable trend attests to the robustness of the proposed procedure.

\par
\begin{figure}[H]
\centering
\includegraphics[angle=-90, width=0.86\textwidth]{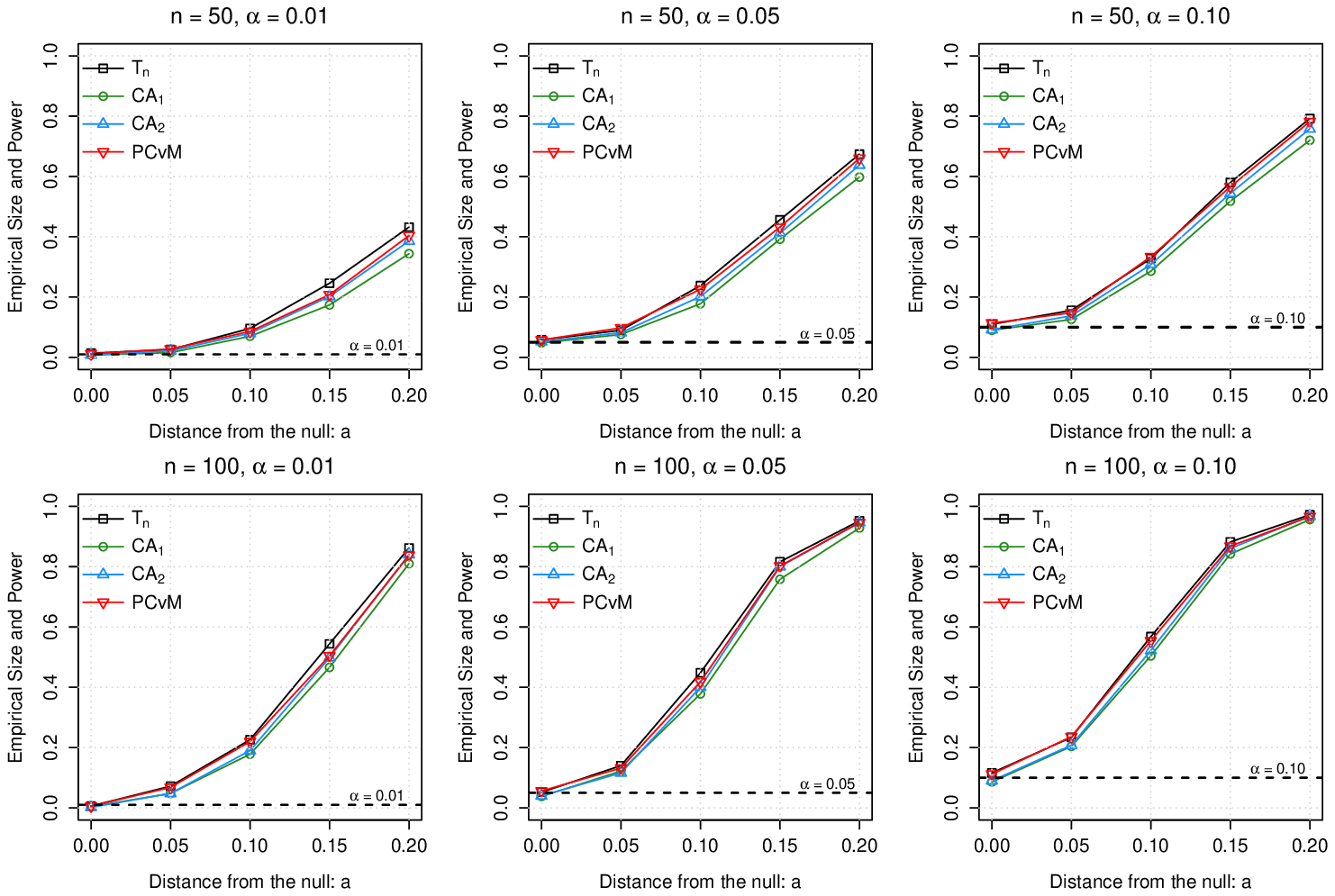}
\caption{Simulation Results for Example \ref{ex1}. The test performance of the proposed test $T_n$, \cite{GarciaPortugues2014}'s tests $PCvM$, and \cite{CuestaAlbertos2019}'s test $CA_1$ and $CA_2$ using \cite{shang&cheng2015inference}'s estimation method at 1\%, 5\% and 10\% significance level with $n = 50, ~100$ and data-driven $p$.}
\label{fig:ex1-2}
\end{figure}

\begin{figure}[H]
\centering
\includegraphics[angle=-90, width=0.86\textwidth]{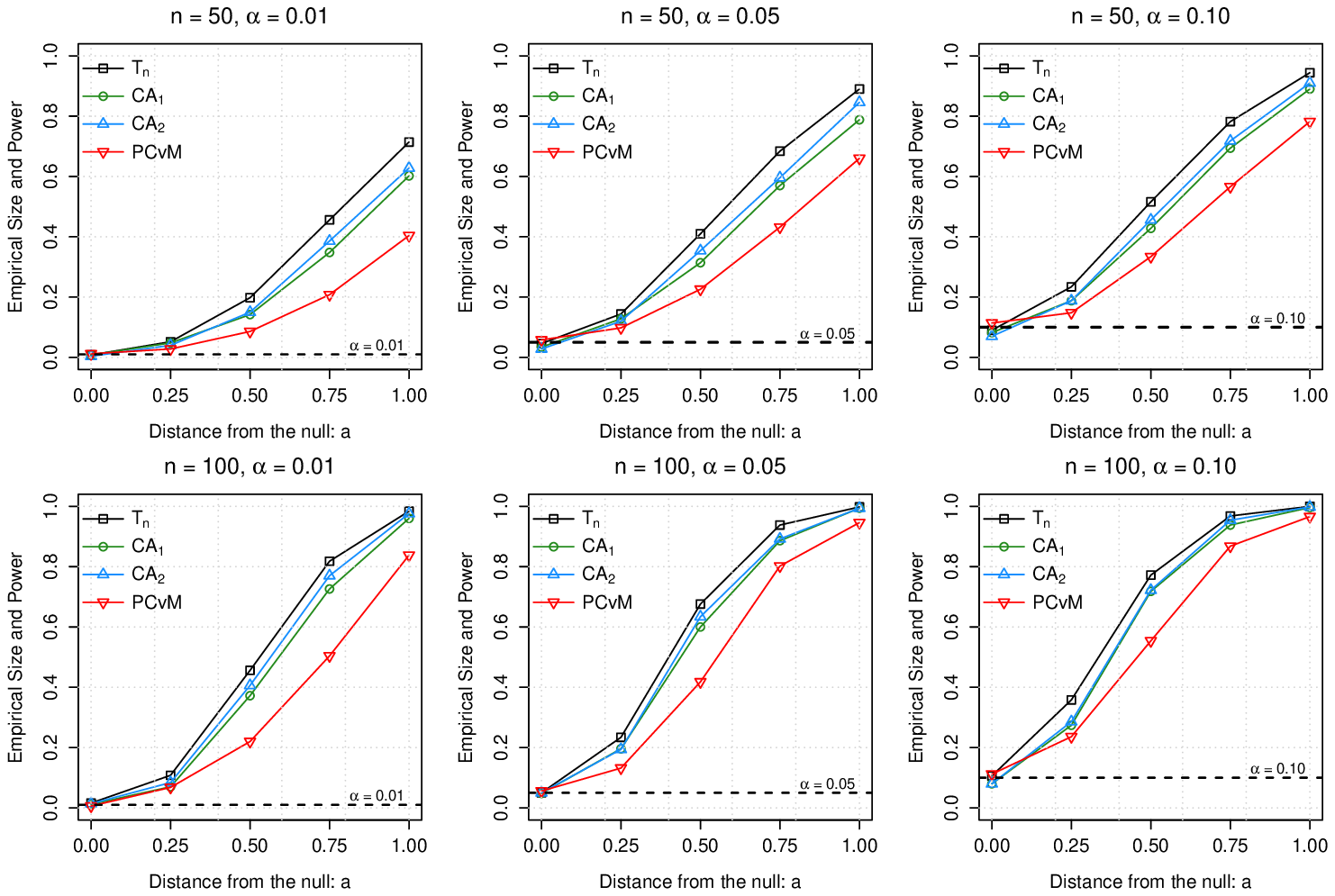}
\caption{Simulation Results for Example \ref{ex2}. The test performance of the proposed test $T_n$, \cite{GarciaPortugues2014}'s tests $PCvM$, and \cite{CuestaAlbertos2019}'s test $CA_1$ and $CA_2$ at 1\%, 5\% and 10\% significance level with $n = 50, ~100$ and data-driven $p$.} 
\label{fig:ex2-1}
\end{figure}

 We present a unified comparison by visualizing in Figures \ref{fig:ex1-2} and \ref{fig:ex2-1} the results corresponding specifically to the \cite{shang&cheng2015inference} estimator from Tables \ref{tab:2} and \ref{tab:3}, respectively. Figure \ref{fig:ex1-2} demonstrates that when the null hypothesis holds with $a=0$, $T_n$ maintains good control of the empirical size. Under the alternatives, its empirical power grows with both the distance from the null $a$ and the sample size $n$, confirming the test's consistency. Furthermore, $T_n$ exhibits greater power at smaller significance levels $\alpha$ and larger values of $a$. Even at $\alpha=0.10$, its performance remains comparable to $PCvM$ and is slightly superior to $CA_1$ and $CA_2$. Turning to Example \ref{ex2}, which involves a functional logistic regression model, Figure \ref{fig:ex2-1} shows that $T_n$ achieves substantially greater power across all simulation conditions. The proposed test demonstrates strong and reliable performance in both examples, confirming its broad applicability for GFLMs and its effectiveness at small significance levels.

\begin{table}[H]
    \centering
    \footnotesize
    \caption{Simulation Results for Example \ref{ex1}: Empirical size and power of the proposed test $T_n$, comparing data-driven $p$ and fixed $p = 5, 10$ selection of the orthonormal bases for sample sizes $n = 50, 100$ at 1\%, 5\%, and 10\% significance level.}
    \label{tab:1}
    
    \begin{tabular}{cc ccc ccc ccc}
        \toprule
        \multirow{2}{*}{$n$} & \multirow{2}{*}{$a$} &
        \multicolumn{3}{c}{1\% Level} & \multicolumn{3}{c}{5\% Level} & \multicolumn{3}{c}{10\% Level} \\
        \cmidrule(lr){3-5} \cmidrule(lr){6-8} \cmidrule(l){9-11}
        & & {data-driven} & {5} & {10} & {data-driven} & {5} & {10} & {data-driven} & {5} & {10} \\
        \midrule
        
         50 & 0 & 0.014 & 0.014 & 0.014 & 0.058 & 0.060 & 0.058 & 0.110 & 0.102 & 0.106 \\
          & 0.05 & 0.026 & 0.032 & 0.030 & 0.090 & 0.094 & 0.092 & 0.156 & 0.150 & 0.152 \\
          & 0.1  & 0.096 & 0.088 & 0.088 & 0.238 & 0.240 & 0.242 & 0.328 & 0.334 & 0.338 \\
          & 0.15 & 0.246 & 0.242 & 0.242 & 0.456 & 0.452 & 0.448 & 0.580 & 0.582 & 0.584 \\
          & 0.2  & 0.432 & 0.434 & 0.440 & 0.674 & 0.674 & 0.672 & 0.792 & 0.786 & 0.786 \\

        \midrule
        \addlinespace[0.1em]
        
         100 & 0 & 0.006 & 0.004 & 0.004 & 0.052 & 0.054 & 0.054 & 0.116 & 0.116 & 0.108 \\
          & 0.05 & 0.072 & 0.070 & 0.068 & 0.140 & 0.132 & 0.132 & 0.234 & 0.238 & 0.238 \\
          & 0.1  & 0.226 & 0.228 & 0.226 & 0.448 & 0.432 & 0.430 & 0.568 & 0.568 & 0.574 \\
          & 0.15 & 0.544 & 0.526 & 0.532 & 0.816 & 0.812 & 0.816 & 0.882 & 0.880 & 0.884 \\
          & 0.2  & 0.862 & 0.858 & 0.862 & 0.952 & 0.948 & 0.950 & 0.972 & 0.972 & 0.972 \\
        
        

        
        
        \bottomrule
    \end{tabular}
\end{table}

\begin{table}[H]
    \centering
	\footnotesize
    \setlength{\tabcolsep}{4pt}
	\caption{Simulation Results for Example \ref{ex1}. Empirical size and power of the proposed test $T_n$, \cite{GarciaPortugues2014}'s tests $PCvM$, and \cite{CuestaAlbertos2019}'s test $CA_1$ and $CA_2$ at 1\%, 5\% and 10\% significance level with $n = 50, ~100$ and data-driven $p$.}
    \label{tab:2}%
    
    \begin{tabular}{ccc cccc cccc cccc}
        \toprule
        \multirow{2}{*}{Est.method} & \multirow{2}{*}{$n$} & \multirow{2}{*}{$a$} & 
        \multicolumn{4}{c}{1\% Level} & \multicolumn{4}{c}{5\% Level} & \multicolumn{4}{c}{10\% Level} \\
        \cmidrule(lr){4-7} \cmidrule(lr){8-11} \cmidrule(l){12-15}
        & & & {$T_n$} & {$CA_1$} & {$CA_2$} & {$PCvM$} & {$T_n$} & {$CA_1$} & {$CA_2$} & {$PCvM$} & {$T_n$} & {$CA_1$} & {$CA_2$} & {$PCvM$} \\
        \midrule
        
        SC & 50 & 0 & 0.014 & 0.008 & 0.008 & 0.012 & 0.058 & 0.048 & 0.052 & 0.058 & 0.110 & 0.090 & 0.092 & 0.114 \\
         &  & 0.05 & 0.026 & 0.016 & 0.022 & 0.028 & 0.090 & 0.076 & 0.082 & 0.098 & 0.156 & 0.126 & 0.138 & 0.148 \\
         &  & 0.1  & 0.096 & 0.070 & 0.080 & 0.086 & 0.238 & 0.178 & 0.202 & 0.226 & 0.328 & 0.286 & 0.308 & 0.334 \\
         &  & 0.15 & 0.246 & 0.174 & 0.202 & 0.208 & 0.456 & 0.392 & 0.412 & 0.432 & 0.580 & 0.518 & 0.544 & 0.566 \\
         &  & 0.2  & 0.432 & 0.344 & 0.386 & 0.404 & 0.674 & 0.598 & 0.638 & 0.660 & 0.792 & 0.720 & 0.758 & 0.782 \\

        \cline{2-15}
        \addlinespace[0.1em]
        
         & 100 & 0 & 0.006 & 0.004 & 0.002 & 0.006 & 0.052 & 0.038 & 0.040 & 0.056 & 0.116 & 0.086 & 0.090 & 0.112 \\
         &  & 0.05 & 0.072 & 0.048 & 0.048 & 0.068 & 0.140 & 0.122 & 0.116 & 0.132 & 0.234 & 0.204 & 0.208 & 0.236 \\
         &  & 0.1  & 0.226 & 0.178 & 0.190 & 0.220 & 0.448 & 0.378 & 0.400 & 0.418 & 0.568 & 0.504 & 0.522 & 0.554 \\
         &  & 0.15 & 0.544 & 0.466 & 0.498 & 0.504 & 0.816 & 0.758 & 0.800 & 0.802 & 0.882 & 0.842 & 0.860 & 0.868 \\
         &  & 0.2  & 0.862 & 0.810 & 0.840 & 0.838 & 0.952 & 0.928 & 0.946 & 0.946 & 0.972 & 0.956 & 0.968 & 0.966 \\
        
        \midrule
        \addlinespace[0.3em]
        
        CAP & 50 & 0 & 0.016 & 0.010 & 0.008 & 0.014 & 0.058 & 0.046 & 0.046 & 0.058 & 0.108 & 0.078 & 0.086 & 0.106 \\
         &  & 0.05 & 0.024 & 0.022 & 0.022 & 0.026 & 0.100 & 0.082 & 0.084 & 0.100 & 0.152 & 0.128 & 0.128 & 0.144 \\
         &  & 0.1  & 0.096 & 0.074 & 0.082 & 0.086 & 0.240 & 0.188 & 0.206 & 0.238 & 0.342 & 0.294 & 0.304 & 0.338 \\
         &  & 0.15 & 0.248 & 0.184 & 0.196 & 0.224 & 0.472 & 0.398 & 0.438 & 0.452 & 0.588 & 0.532 & 0.552 & 0.578 \\
         &  & 0.2  & 0.446 & 0.360 & 0.398 & 0.416 & 0.688 & 0.614 & 0.664 & 0.672 & 0.798 & 0.724 & 0.764 & 0.796 \\

        \cline{2-15}
        \addlinespace[0.1em]
        
         & 100 & 0 & 0.010 & 0.002 & 0.004 & 0.010 & 0.050 & 0.040 & 0.036 & 0.052 & 0.106 & 0.080 & 0.092 & 0.108 \\
         &  & 0.05 & 0.068 & 0.050 & 0.048 & 0.068 & 0.144 & 0.118 & 0.122 & 0.134 & 0.242 & 0.208 & 0.206 & 0.232 \\
         &  & 0.1  & 0.230 & 0.182 & 0.200 & 0.214 & 0.450 & 0.382 & 0.396 & 0.410 & 0.566 & 0.496 & 0.524 & 0.556 \\
         &  & 0.15 & 0.540 & 0.460 & 0.510 & 0.514 & 0.822 & 0.756 & 0.800 & 0.800 & 0.878 & 0.848 & 0.860 & 0.874 \\
         &  & 0.2  & 0.868 & 0.816 & 0.852 & 0.850 & 0.952 & 0.922 & 0.946 & 0.942 & 0.974 & 0.954 & 0.968 & 0.968 \\
        
        \bottomrule
    \end{tabular}
\end{table}

\begin{table}[H]
    \centering
	\footnotesize
	\caption{Simulation Results for Example \ref{ex2}. Empirical size and power of the proposed test $T_n$, \cite{GarciaPortugues2014}'s tests $PCvM$, and \cite{CuestaAlbertos2019}'s test $CA_1$ and $CA_2$ at 1\%, 5\% and 10\% significance level with $n = 50, ~100$ and data-driven $p$.}
    \label{tab:3}%
    \begin{tabular}{cc cccc cccc cccc}
        \toprule
        \multirow{2}{*}{$n$} & \multirow{2}{*}{$a$} & 
        \multicolumn{4}{c}{1\% Level} & \multicolumn{4}{c}{5\% Level} & \multicolumn{4}{c}{10\% Level} \\
        \cmidrule(lr){3-6} \cmidrule(lr){7-10} \cmidrule(l){11-14}
        & & {$T_n$} & {$CA_1$} & {$CA_2$} & {$PCvM$} & {$T_n$} & {$CA_1$} & {$CA_2$} & {$PCvM$} & {$T_n$} & {$CA_1$} & {$CA_2$} & {$PCvM$} \\
        \midrule
        
        50 & 0 & 0.008 & 0.008 & 0.004 & 0.008 & 0.046 & 0.034 & 0.028 & 0.046 & 0.090 & 0.082 & 0.070 & 0.088 \\
         & 0.25 & 0.052 & 0.048 & 0.040 & 0.050 & 0.144 & 0.128 & 0.120 & 0.146 & 0.234 & 0.188 & 0.188 & 0.222 \\
         & 0.5  & 0.198 & 0.142 & 0.150 & 0.182 & 0.410 & 0.314 & 0.354 & 0.384 & 0.516 & 0.428 & 0.456 & 0.494 \\
         & 0.75 & 0.456 & 0.348 & 0.386 & 0.410 & 0.684 & 0.570 & 0.596 & 0.654 & 0.782 & 0.694 & 0.718 & 0.752 \\
         & 1    & 0.714 & 0.602 & 0.628 & 0.694 & 0.890 & 0.788 & 0.846 & 0.870 & 0.944 & 0.890 & 0.910 & 0.930 \\
        
        \midrule
        \addlinespace[0.1em]
        
         100 & 0 & 0.016 & 0.012 & 0.014 & 0.014 & 0.050 & 0.048 & 0.050 & 0.054 & 0.106 & 0.080 & 0.080 & 0.090 \\
          & 0.25 & 0.108 & 0.070 & 0.084 & 0.100 & 0.234 & 0.196 & 0.194 & 0.236 & 0.358 & 0.274 & 0.286 & 0.332 \\
          & 0.5  & 0.456 & 0.372 & 0.406 & 0.432 & 0.676 & 0.600 & 0.634 & 0.666 & 0.772 & 0.718 & 0.722 & 0.752 \\
          & 0.75 & 0.818 & 0.726 & 0.770 & 0.806 & 0.938 & 0.886 & 0.892 & 0.934 & 0.968 & 0.938 & 0.954 & 0.962 \\
          & 1    & 0.984 & 0.960 & 0.976 & 0.978 & 0.998 & 0.994 & 0.994 & 0.996 & 1.000 & 0.996 & 0.998 & 1.000 \\
        
        \bottomrule
    \end{tabular}
\end{table}

\section{Discussion}\label{conclusion}

This paper develops a novel goodness-of-fit test for GFLMs, a domain where model checking procedures are still relatively scarce. A fundamental challenge in this area arises from the infinite-dimensional nature of functional predictors, which often renders traditional testing methodologies computationally intractable or theoretically difficult. To overcome this difficulty, our method introduces a projection averaging framework that culminates in a $U$-statistic. This formulation effectively transforms the intractable problem of integrating over all projection directions into a tractable analytical form. A central theoretical contribution of this work is the derivation of the asymptotic null distribution for this $U$-statistic. We establish that our test is consistent under the alternatives, ensuring its ability to detect a wide range of model misspecifications. The finite-sample validity of these theoretical results was confirmed by the simulation studies in Section 4.

The developed framework warrants further investigation in several promising directions. An extension involves adapting our methodology to broader classes of functional models, such as functional additive models or semi-parametric models with an unknown link function. Moreover, the core principle of projection averaging presents a promising approach for goodness-of-fit tests for high-dimensional regression. This problem is fundamentally distinct from existing applications of projection averaging methods, such as the two-sample test in \cite{Kim2020}, which does not involve parameter estimation. A crucial challenge for a goodness-of-fit test is to rigorously account for the influence of the preliminary penalized estimators on which the test residuals depend, thereby constituting a significant area for future research.

\section*{Acknowledgments}
This work was supported by the National Natural Science Foundation of China (12101055) and the Guangdong Basic and Applied Basic Research Foundation (2024A1515011453).

\bibliography{vcms}
\bibliographystyle{elsarticle-harv}\biboptions{authoryear}

\end{document}